\documentclass[
 reprint,
 amsmath,amssymb,
 aps,
]{revtex4-2}

\usepackage{graphicx}
\usepackage{dcolumn}
\usepackage{bm}
\usepackage{braket}
\usepackage{amsfonts}
\usepackage{dsfont}
\usepackage{xcolor}
\usepackage{mathtools}
\usepackage{hyperref}

\newcommand{\Nin}{N_{\text{in}}}
\newcommand{\Nout}{N_{\text{out}}}

\begin{document}

\title{Enhanced Squeezing and Faster Metrology from Layered Quantum Neural Networks}

\author{Nickholas Gutierrez}
\author{Rodrigo Araiza Bravo}
\author{Susanne Yelin}
\affiliation{Harvard University, 60 Oxford St., Cambridge, MA 02138}

\date{\today}

\begin{abstract}
Spin squeezing is a powerful resource for quantum metrology, and recent hardware platforms based on interacting qubits provide multiple possible architectures to generate and reverse squeezing during a sensing protocol. In this work, we compare the sensing performance of three such architectures—quantum reservoir computers (QRCs), quantum perceptrons, and multi-layer quantum neural networks (QNNs)—when used as squeezing-based field sensors. For all models, we consider a standard metrological sequence consisting of coherent-spin preparation, one-axis-twisting dynamics, field encoding via a weak rotation, time-reversal, and collective readout. We show that a single quantum perceptron generates the same optimal sensitivity as a QRC, but in the perturbative regime it benefits from accelerated squeezing due to steering by the output qubit. Stacking perceptrons into a QNN further amplifies this effect: in a 2-layer QNN with $\Nin$ input and $\Nout$ output qubits, the optimal squeezing time is reduced by a factor of $\Nout$, while the achievable phase sensitivity remains Heisenberg-limited, $\Delta\phi \sim 1/(\Nin+\Nout)$. Moreover, if the layers are used sequentially—first using the outputs to squeeze the inputs and then the inputs to squeeze the outputs—the two contributions to the response add constructively. This yields a $\sqrt{2}$ enhancement in sensitivity over a QRC when $\Nin=\Nout$ and requires shorter total squeezing time. Generalizing to $L$ layers, we show that the metrological gain scales as $\sqrt{L}$ while the required squeezing time decreases as $1/N_l$, where $N_l$ is the number of qubits per layer. Our results demonstrate that the structure of quantum neural networks can be exploited not only for computation, but also to engineer faster and more sensitive squeezing-based quantum sensors.
\end{abstract}

\maketitle

\section{Introduction}
Spin squeezing has emerged as a leading pathway toward quantum-enhanced metrology because it enables sensitivities below the standard quantum limit (SQL), eventually reaching the Heisenberg limit when correlations are engineered across $N$ qubits \cite{davis2016approaching,wineland1994squeezed,sorensen2001entanglement}. The one-axis–twisting (OAT) Hamiltonian, $H_{\text{OAT}} = \chi S_z^2$, remains one of the most experimentally promising mechanisms to generate the entanglement required for such enhancement \cite{kitagawa1993squeezed}. Beyond its conceptual simplicity, the OAT mechanism has been realized across platforms including trapped ions, neutral atoms, cavity QED, and superconducting qubits \cite{bohnet2016quantum,riedel2010atom,gross2010nonlinear,cox2016spin,hosten2016measurement}. This makes squeezing-based sensing not only theoretically attractive but practically relevant in near-term quantum hardware.

Historically, squeezing analyses focus on a \emph{single} monolithic ensemble of $N$ qubits evolving under OAT before and after interrogation by a signal field. Yet current quantum architectures do not always present qubits as a single homogeneous entity. Instead, near-term devices often adopt structured, layered, or modular layouts inspired by the architectures of computational quantum models such as quantum reservoir computers (QRC), quantum perceptrons, and quantum neural networks (QNN). This raises a crucial question for metrology: \emph{does the computational architecture of the qubit connectivity influence the achievable sensing performance, even when the underlying entangling Hamiltonian is the same?}

From a metrological perspective, not all architectures are equivalent. A QRC applies OAT squeezing simultaneously across all qubits. A quantum perceptron, in contrast, squeezes only a subset of the system (the inputs) via an auxiliary node (the output), and a QNN stacks perceptrons such that input and output blocks can squeeze one another. These differences lead not only to different total squeezing resources but also to different \emph{temporal ordering} of squeezing and un-squeezing operations—something that matters only because squeezing is exponentially sensitive to angle misalignment \cite{cox2016spin,wineland1994squeezed}.

In other words, even if every architecture reaches the Heisenberg limit asymptotically, the \emph{squeezing time} and the \emph{prefactor of the sensitivity} depend on how the system is structured. Time resources can become the dominant bottleneck in real hardware, where squeezing must be generated in a regime that avoids decoherence, leakage, and breakdown of the perturbative approximations that motivated the effective Hamiltonians in the first place \cite{hosten2016measurement,cox2016spin}. Thus, architectures that accelerate squeezing without sacrificing sensitivity are of practical importance.

This work compares the squeezing-based sensing performance across three architectures:
\begin{enumerate}
    \item a Quantum Reservoir Computer (QRC),
    \item a Quantum Perceptron,
    \item a Quantum Neural Network (QNN).
\end{enumerate}
All three ultimately attain Heisenberg scaling in $N$, but we show that the QNN performs the squeezing in a fraction of the time relative to a QRC and, crucially, that a \textbf{2-layer QNN achieves both}:
\begin{itemize}
    \item a \emph{$1/\Nout$ reduction in squeezing time} due to accelerated effective OAT dynamics, and
    \item a \emph{$\sqrt{2}$ improvement in sensitivity} compared to a QRC with the same total number of qubits.
\end{itemize}
The enhancement has a simple physical origin: squeezing is first generated on one block of the system while the other acts as an accelerator, and then the roles are reversed. When squeezing is performed sequentially rather than simultaneously, the contributions to $\partial_{\phi}\langle S_y\rangle$ \emph{add} rather than \emph{coalesce}, producing a measurable prefactor improvement while maintaining Heisenberg scaling. This observation generalizes to $L$ layers, where the sensitivity enhancement grows as $\sqrt{L}$.

The remainder of the paper shows this effect using direct calculations of the squeezed sensing response and compares the QRC, perceptron, and QNN architectures under equivalent OAT resources.

\begin{figure*}[t]
    \centering
    \includegraphics[height=0.48\textheight]{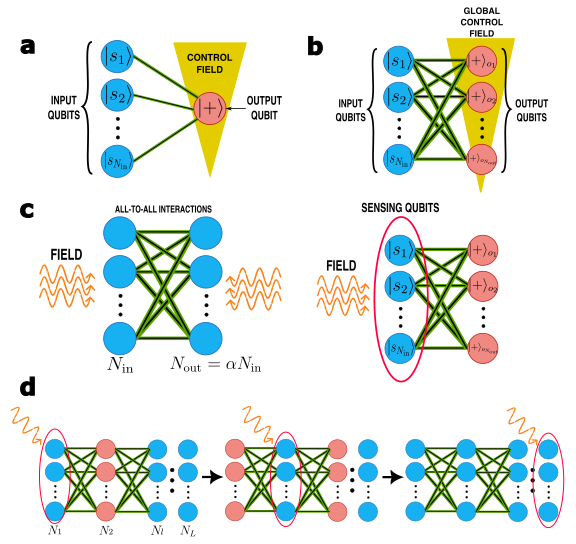}
    \caption{
Using quantum perceptrons for quantum metrology. 
(a) A single quantum perceptron composed of $N$ input qubits interacting via $ZZ$ couplings with a single output qubit, which is prepared in the $\ket{+}$ state during time reversal. Only the output qubit is directly controlled. 
(b) Stacking perceptrons such that each qubit in the rightmost layer interacts via $ZZ$ couplings with every qubit in the leftmost layer, with global control applied to the rightmost layer. 
(c) Two metrological schemes using multiple perceptrons: the left configuration treats the entire network with all-to-all interactions as a single sensor, while the right configuration senses only with the leftmost layer while qubits in the second layer are driven out. 
(d) Sensing protocol for a QNN with $L$ layers. One senses with the first layer and sequentially propagates the sensing operation through each layer. Neighboring layers drive the one-axis-twisting and time-reversal dynamics. This protocol enables all qubits to participate as sensors, achieving both maximal precision and a reduced optimal squeezing time.
}
    \label{fig:fullPerceptronMetrology}
\end{figure*}

\section{Sensing with a Single Quantum Perceptron}

We begin by analyzing the simplest nontrivial structured architecture: a quantum perceptron consisting of $\Nin$ input qubits coupled to a single output qubit. Although perceptrons are usually motivated by quantum machine-learning applications, here they provide a minimal model in which architectural asymmetry produces meaningful changes to squeezing dynamics. The Hamiltonian is
\begin{equation}
    H_P = J\sum_{i=1}^{\Nin}Z_iZ_{\text{out}},
\end{equation}
a star-graph interaction pattern in which the output qubit serves as a mediator for all input–input correlations. This geometry concentrates all nonlinear resources onto the output node; however, when the output is strongly driven, it behaves almost classically and feeds back an effective interaction among the inputs.

Applying a large Rabi field $\Omega$ along $x$ on the output qubit, with $\Omega \gg J$, generates an effective interaction through second-order processes. After averaging in the toggling frame, the resulting Hamiltonian becomes \cite{bravo2022universal}
\begin{equation}
    H_{\text{eff}} = \frac{J^2}{2\Omega}\sum_{i,j=1}^{\Nin}Z_iZ_j \otimes X_{\text{out}}
    = \chi (S_z^{\text{in}})^2 \otimes X_{\text{out}}.
\end{equation}
The physical meaning is that although the inputs do not couple directly, they “borrow’’ nonlinear interactions via the output qubit. The output itself remains essentially unsqueezed, but its stabilized $x$-polarization ensures that it faithfully mediates a one-axis-twisting (OAT) term for the input block. Thus the perceptron reproduces the full OAT Hamiltonian on $\Nin$ qubits despite using only a single nonlinear mediator.

A standard squeezing-based metrological sequence—coherent-spin preparation, forward squeezing under $H_{\text{eff}}$, weak rotation by $\phi$, backward evolution, and measurement of $S_y^{\text{in}}$—yields the same Heisenberg-limited scaling as in an all-to-all OAT system with $N=\Nin$. The optimal squeezing strength is identical:
\begin{equation}
    Q_{\text{opt}}^{P}= \Nin\text{arccot}(\sqrt{\Nin-2}), 
    \qquad 
    \theta^{\text{P}}_{\text{opt}}=\text{arccot}(\sqrt{\Nin-2}).
\end{equation}
Thus a single perceptron offers no fundamental metrological advantage over a conventional ensemble of $N$ qubits. However, the important architectural feature is that the squeezing rate is enhanced in the perturbative regime because the output qubit continuously steers the inputs. This effect becomes significantly amplified once multiple output qubits are introduced.

\section{Sensing with a Quantum Neural Network}

We now extend the perceptron model to a quantum neural network (QNN) with $\Nin$ input qubits and $\Nout$ output qubits. In this multilayer architecture, each output qubit behaves like an independent mediator, producing many parallel second-order pathways that generate the effective nonlinear interaction. The Hamiltonian is
\begin{equation}
    H_{\text{QNN}} = \sum_{j=1}^{\Nout}H_P^{j} 
    = J\sum_{i=1}^{\Nin}\sum_{j=1}^{\Nout} Z_iZ_j.
\end{equation}
When each output qubit is driven strongly along $x$ and prepared in a coherent spin state, the combined effect of all output qubits enhances the OAT interaction experienced by the inputs. Explicitly,
\begin{equation}
    e^{-i\chi (S_z^{\text{in}})^2\otimes\sum_j X_j}
    = e^{-i\Nout\chi (S_z^{\text{in}})^2}
    = U_{\text{in}}^{\Nout}.
\end{equation}

The interpretation is that $\Nout$ steering qubits each contribute additively to the nonlinear twisting rate, so the input block undergoes OAT evolution \emph{N$_\text{out}$ times faster} than in the single-perceptron architecture. This does not alter the functional form of the metrological response; instead it compresses the time required to reach the optimal squeezing point. As a result, the metrological sensitivity remains Heisenberg-limited in the total number of qubits, but the time to generate optimal squeezing is reduced by a factor of $\Nout$. The optimal squeezing angle therefore becomes
\begin{equation}
    \theta^{\text{QNN}}_{\text{opt}} 
    = \frac{Q_{\text{opt}}}{\Nout \Nin} 
    \approx \frac{1}{\Nout\sqrt{\Nin}},
\end{equation}
demonstrating a purely architectural enhancement in the temporal resources required for metrology. This accelerated squeezing mechanism forms the basis for the more dramatic improvements possible in the two-layer QNN.

\section{Sensitivity Improvement with a 2-Layer QNN}

A single perceptron accelerates squeezing but does not change the fundamental sensitivity. In contrast, a 2-layer QNN introduces an additional dynamical feature: each layer can squeeze the other. When squeezing is generated sequentially rather than simultaneously, the metrological contributions add at first order in $\phi$. This additive structure produces a genuine sensitivity enhancement even when the total qubit count matches that of a QRC.

Consider a 2-layer QNN with $\Nin$ input and $\Nout$ output qubits. First, the outputs squeeze the inputs under accelerated OAT dynamics; second, the inputs squeeze the outputs under a symmetric operation. After imprinting a weak field and reversing both stages, the resulting state is
\begin{equation}
    |\psi\rangle = |\text{CSS}'\rangle_{\text{all}} 
    -i\phi \!\left[
    \tilde{U}_{\text{in}}^\dagger S_y^{\text{in}}\tilde{U}_{\text{in}}
    +
    \tilde{U}_{\text{out}}^\dagger S_y^{\text{out}}\tilde{U}_{\text{out}}
    \right]
    |\text{CSS}'\rangle_{\text{all}} 
    + O(\phi^2),
\end{equation}
which contains two independent linear-response contributions. Since the two blocks are initially uncorrelated, their uncertainties combine as
\begin{equation}
    \Delta^{\phi=0} S_y = \sqrt{\Nin/2+\Nout/2}.
\end{equation}

The corresponding derivatives also add:
\begin{align}
    D_{\text{in}}(\theta) &= 
    \frac{\Nin}{2}(\Nin-1)
    \sin\!\left(\Nout\theta\right)
    \cos^{\Nin-2}\!\left(\Nout\theta\right),\\
    D_{\text{out}}(\theta) &= 
    \frac{\Nout}{2}(\Nout-1)
    \sin\!\left(\Nin\theta\right)
    \cos^{\Nout-2}\!\left(\Nin\theta\right),
\end{align}
so that the net response is
\begin{equation}
    \partial_\phi \langle S_y\rangle^{\phi=0} 
    = D_{\text{in}}(\theta)+D_{\text{out}}(\theta).
\end{equation}

For equal layer sizes $\Nin=\Nout=N_l$, the optimal point satisfies $N_l\theta \approx 1/\sqrt{N_l}$, yielding
\begin{equation}
    \partial_\phi \langle S_y\rangle^{\phi=0}
    \approx N_l^{3/2},
\end{equation}
while a QRC with $N=2N_l$ qubits gives
\begin{equation}
    \partial_\phi \langle S_y\rangle^{\phi=0} 
    \approx \sqrt{2}\,N_l^{3/2}.
\end{equation}
Thus the 2-layer QNN achieves a \emph{$\sqrt{2}$ improvement in sensitivity} relative to a QRC of identical size. The mechanism is architectural and dynamical: sequential squeezing produces additive metrological derivatives, whereas simultaneous squeezing collapses all structure into a single contribution. Figure~\ref{fig:Comparison} illustrates this enhancement.

\section{Sensitivity Analysis for Multilayer QNNs}
We now expand the derivation of the sensitivity for the multilayer QNN and make explicit each of the steps leading to the summary scaling described above.

\begin{figure*}[t]
    \centering
    \includegraphics[width=\linewidth]{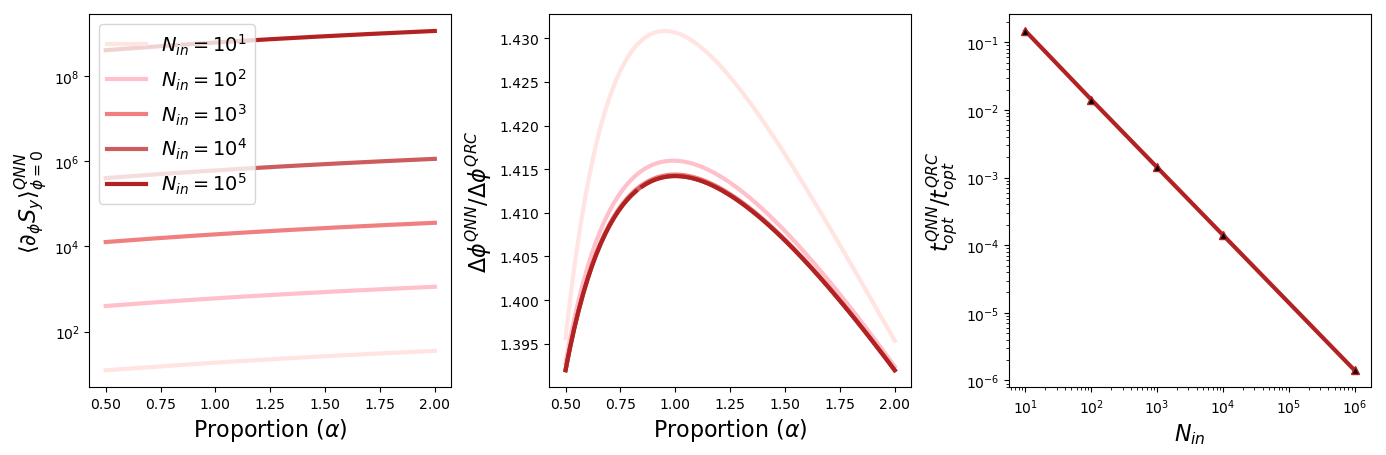}
    \caption{
        Comparison of squeezing dynamics and sensitivity across the QRC, perceptron, and QNN architectures.
    }
    \label{fig:Comparison}
\end{figure*}

For a 2-layer QNN with $\Nin$ and $\Nout = \alpha \Nin$, each block can be squeezed by the other. When the outputs squeeze the inputs, the relevant contribution to the derivative is
\begin{equation}
D_{\text{in}}(\theta) = \frac{\Nin}{2}(\Nin-1)\sin(\Nout\theta)\cos^{\Nin-2}(\Nout\theta),
\end{equation}
where the argument $\Nout\theta$ appears because squeezing is accelerated by a factor of $\Nout$. Conversely, when the inputs squeeze the outputs,
\begin{equation}
D_{\text{out}}(\theta) = \frac{\Nout}{2}(\Nout-1)\sin(\Nin\theta)\cos^{\Nout-2}(\Nin\theta).
\end{equation}
Because these contributions arise from two distinct squeezing-and-unsqueezing stages, they \textbf{add coherently} at first order in $\phi$:
\begin{equation}
\partial_{\phi}\langle S_y\rangle^{\phi=0} = D_{\text{in}}(\theta) + D_{\text{out}}(\theta).
\end{equation}
This comparison is shown explicitly in Fig.~\ref{fig:Comparison}. This is the key difference from the QRC, where all qubits sense simultaneously and the squeezing dynamics do not decompose into additive contributions.

To reveal the scaling, we analyze the optimal point. Specializing to $\alpha = 1$, i.e., $\Nout = \Nin \equiv N_l$, we have
\begin{equation}
D_{\text{in}} = D_{\text{out}} = \frac{N_l}{2}(N_l-1)\sin(N_l\theta)\cos^{N_l-2}(N_l\theta).
\end{equation}
Thus
\begin{equation}
\partial_{\phi}\langle S_y\rangle^{\phi=0} = N_l(N_l-1)\sin(N_l\theta)\cos^{N_l-2}(N_l\theta).
\end{equation}
The maximum occurs when
\begin{equation}
N_l\theta = \text{arccot}\!\sqrt{N_l - 2} \approx \frac{1}{\sqrt{N_l}}  \quad \text{for large } N_l.
\end{equation}
Substituting this value yields the asymptotic scaling
\begin{equation}
\partial_{\phi}\langle S_y\rangle^{\phi=0} \approx N_l^{3/2}.
\end{equation}

A QRC with the same total number of qubits $N = 2N_l$ gives instead
\begin{equation}
\partial_{\phi}\langle S_y\rangle^{\phi=0} \approx \sqrt{2}\,N_l^{3/2},
\end{equation}
demonstrating that the QNN enjoys a \textbf{$\sqrt{2}$ advantage in sensitivity} while retaining full Heisenberg scaling.

We now generalize to $L$ layers with $N_l$ qubits per layer and $\alpha = 1$. Each interior layer (those not at the edges) is squeezed twice—once by the layer to the left and once by the layer to the right. End layers are squeezed once. Thus the number of effective squeezing events is
\begin{equation}
\#\text{events} = 2(L-2) + 2 = 2(L-1),
\end{equation}
and the total derivative becomes
\begin{equation}
\partial_{\phi}\langle S_y\rangle^{\phi=0} \approx (L-1)\, N_l^{3/2}.
\end{equation}
For a QRC with $N = L N_l$ qubits, the corresponding scaling is
\begin{equation}
\partial_{\phi}\langle S_y\rangle^{\phi=0} \approx (L N_l)^{3/2} = L^{3/2} N_l^{3/2}.
\end{equation}
Dividing the two sensitivities gives the relative advantage,
\begin{equation}
\frac{\Delta\phi^{\text{QNN}}(L)}{\Delta\phi^{\text{QRC}}(L)} = \frac{\sqrt{L}}{1 - \frac{N_l + L}{N_l L + 1}} \approx \frac{1}{\sqrt{L}} \quad \text{for } L \ll N_l,
\end{equation}
revealing a \textbf{$\sqrt{L}$ enhancement} in sensitivity for the multilayer QNN over an architecture that squeezes all qubits simultaneously.

In addition, the time to generate optimal squeezing in each QNN layer scales as $\theta_{\text{opt}} \approx 1/\sqrt{N_l}$ and is implemented using accelerated dynamics proportional to the number of qubits steering the squeezing. As every layer is steered by $N_l$ qubits, the squeezing time is reduced by \textbf{$1/N_l$} compared to the QRC. Thus, multilayer QNNs achieve both \textbf{faster squeezing} and \textbf{higher sensitivity}.

\section{Discussion}
The results presented in this work show that the architecture of an interacting-qubit system plays a decisive role in squeezing-based metrology, even when the underlying entangling mechanism—the one-axis-twisting Hamiltonian—is kept fixed. In conventional analyses, metrological gain is treated as a function of particle number and squeezing strength alone. However, near-term quantum devices increasingly rely on structured layouts inspired by quantum machine-learning models, and our findings demonstrate that this structure can be exploited as a resource for sensing.

A quantum reservoir computer (QRC) realizes the standard OAT protocol by squeezing all qubits simultaneously, achieving Heisenberg-limited sensitivity with respect to the total qubit count. A single quantum perceptron replaces the monolithic interaction with an asymmetric layout in which a single output qubit steers the dynamics of an input block. Importantly, we find that the perceptron retains the same fundamental metrological sensitivity as a QRC but benefits from faster squeezing in the perturbative regime due to the explicit asymmetry introduced by the output node. This effect becomes amplified when perceptrons are arranged into a quantum neural network (QNN): a 2-layer architecture achieves both a $1/\Nout$ reduction in optimal squeezing time and a measurable prefactor advantage in phase sensitivity, amounting to a $\sqrt{2}$ improvement compared to a QRC of the same total size when $\Nin = \Nout$.

The physical mechanism behind this gain is simple: the input and output layers squeeze one another sequentially rather than simultaneously. Because the metrological response depends on the derivative of the collective spin expectation after time reversal, sequential squeezing stages contribute additively at first order in the signal angle. In contrast, simultaneous squeezing collapses the response into a single contribution. Extending this logic to an $L$-layer QNN, we find that the contributions sum constructively across layers, yielding a $\sqrt{L}$ enhancement in sensitivity and a squeezing time that decreases as $1/N_l$ for $N_l$ qubits per layer. Thus, multilayer QNNs can both reduce the total time required for optimal squeezing and increase signal response—two of the most important bottlenecks for practical quantum sensors operating in decoherence-limited regimes.

Several implications follow. First, the metrological performance of a hardware platform depends not only on the strength of its native interactions, but also on how qubits are grouped and ordered in time during entangling operations. This suggests that architectures originally developed for quantum information processing may unlock sensing advantages without requiring new primitives. Second, even in platforms where full all-to-all connectivity is available, using it in a layered fashion may be more advantageous than applying it homogeneously. Third, because the improvements observed here stem from temporal ordering and steering effects rather than from more exotic protocols (e.g., nonlinear control or non-Gaussian measurements), implementations should be accessible on current neutral-atom, trapped-ion, superconducting-qubit, and cavity-QED hardware.

Looking ahead, several directions remain open. The present analysis assumes ideal time-reversal, negligible decoherence, and a weak-field linear-response regime. Understanding whether architectural advantages persist under dissipation, leakage, or in the presence of control noise is an important next step. Moreover, it would be interesting to compare QNN-based squeezing with alternative mechanisms for fast entanglement generation, including programmable Floquet sequences and disordered or scrambling Hamiltonians. Finally, an open question is whether QNN architectures can be co-designed for dual functionality—performing both efficient sensing and computation within a single framework.

In summary, this work shows that squeezing-based quantum metrology need not be restricted to monolithic ensembles: structured connectivity can be used to engineer faster and more sensitive quantum sensors. Quantum neural networks, far from being solely tools for computation, naturally implement time-optimized and sensitivity-enhanced squeezing protocols. As quantum control hardware matures, harnessing architectural features—not only Hamiltonians—may become a central design principle for quantum sensors.

\section*{Acknowledgments}
This work was supported by the Harvard Quantum Initiative post-baccalaureate research fellowship. The authors are grateful to the Yelin group for a stimulating research environment and ongoing conversations about quantum sensing.

\bibliographystyle{apsrev4-2}
\bibliography{references}

\end{document}